# Presence of a Monoclinic (Pm) Phase in the Morphotropic Phase Boundary Region of Multiferroic (1-x)Bi(Ni$_{1/2}$Ti$_{1/2}$)O$_3$-xPbTiO$_3$ Solid Solution: A Rietveld Study

Rishikesh Pandey and Akhilesh Kumar Singh*

School of Materials Science & Technology, Indian Institute of Technology (Banaras Hindu University) Varanasi, India-221005

## Abstract

We present here the results of structural studies on multiferroic (1-x)Bi(Ni$_{1/2}$Ti$_{1/2}$)O$_3$-xPbTiO$_3$ solid solution using Rietveld analysis on powder x-ray diffraction data in the composition range $0.35 \leq x \leq 0.55$. The stability region of various crystallographic phases at room temperature for (1-x)Bi(Ni$_{1/2}$Ti$_{1/2}$)O$_3$-xPbTiO$_3$ is determined precisely. Structural transformation from pseudo-cubic ($x \leq 0.40$) to tetragonal ($x \geq 0.50$) phase is observed via phase coexistence region demarcating the morphotropic phase boundary. The morphotropic phase boundary region consists of coexisting tetragonal and monoclinic structures with space group P4mm and Pm, respectively, stable in composition range $0.41 \leq x \leq 0.49$ as confirmed by Rietveld analysis. The results of Rietveld analysis completely rule out the coexistence of rhombohedral and tetragonal phases in the morphotropic phase boundary region reported by earlier workers. A comparison between the bond lengths for "*B*-site cations- oxygen anions" obtained after Rietveld refinement, with the bond length calculated using Shannon-Prewitt ionic radii, reveals the ionic nature of *B-O* (Ni/Ti-O) bonds for the cubic phase and partial covalent character for the other crystallographic phases.

\* Corresponding author

E-mail address: akhilesh_bhu@yahoo.com, aksingh.mst@itbhu.ac.in

## I. INTRODUCTION

Recently, there has been a great interest in Bi-based perovskite solid solutions due to potential applications in commercial multifunctional devices [1] and as a substitute of $(PbZr_xTi_{1-x})O_3$ (PZT) and $(1-x)Pb(Mg_{1/2}Nb_{1/2})O_3$-$xPbTiO_3$ (PMN-PT) piezoceramics, which are widely used in sensor and actuator applications [2]. The presence of morphotropic phase boundary (MPB) in these solid solutions is an essential requirement for the maximization of electromechanical and dielectric properties. In the phase diagram of the Pb-based systems such as PZT [3], PMN-PT [4-5], $(1-x)Pb(Zn_{1/3}Nb_{2/3})O_3$-$xPbTiO_3$ (PZN-PT) [6], $(1-x)Pb(Ni_{1/3}Nb_{2/3})O_3$-$xPbTiO_3$ [7] etc, the MPB separates the stability fields of the tetragonal and monoclinic (or pseudocubic) phases [3,8]. The intermediate monoclinic phase in the morphotropic phase boundary (MPB) region is believed to provide the lowest energy path for the polarization rotation, leading to maximization of physical properties [9]. Recently, Bi-based multiferroic systems such as $BiFeO_3$ [10], $Bi(Fe_{0.5}Cr_{0.5})O_3$ [11], $BiMnO_3$ [12-13] etc. have been reported to form the continuous solid solutions with $PbTiO_3$. In multiferroics, more than one ferroic orders such as ferro/antiferroelectricity, ferro/antiferromagnetism and ferroelasticity coexist and may be coupled, which provide extra degree of freedom for designing novel devices [14-15]. It is reported recently that the phase diagram of these multiferroic systems exhibit the phase field of tetragonal and rhombohedral phases separated by an intermediate monoclinic phase [10]. Recently, a new multiferroic solid solution $(1-x)Bi(Ni_{1/2}Ti_{1/2})O_3$-$xPbTiO_3$ (BNT-PT) has been investigated with excellent physical properties and competitive electromechanical coupling coefficient for potential applications in sensors, actuators and memory devices [16]. However, synthesis of its one of the component BNT requires high pressure (~5-6 GPa) as well as high temperature (~ 1000 $^0$C) [17-18] similar to other Bi-based perovskite systems such as $Bi(Zn_{1/2}Ti_{1/2})O_3$ (BZT), $Bi(Mg_{1/2}Zr_{1/2})O_3$, $Bi(Mg_{1/2}Ti_{1/2})O_3$ and $Bi_2MnNiO_6$ [19-23]. Phase pure multiferroic systems at ambient temperature and pressure may be obtained by formation of their solid solutions with other stable perovskites. The crystal and magnetic structure of the stabilized system may be different from the initial components after the formation of solid solution. Following this route, phase pure perovskite form of BNT has been stabilized at ambient pressure by forming its solid solution with other ceramics such as $PbTiO_3$ [24], $Pb_{(1-x)}Sr_xTiO_3$ [25], $BaTiO_3$, $SrTiO_3$ etc. The structure of BNT is double perovskite with orthorhombic space group $Pn2_1a$ obtained by the repetition of monoclinic unit cells [17-18]. However, the structure of BNT entirely changes after the solid solution formation with $PbTiO_3$ at room temperature. The BNT-PT solid solution shows a phase transition from pseudocubic to tetragonal with increasing PT concentration. The MPB is reported to exist around the composition x=0.46 and the structure is reported to be coexistence of rhombohedral and tetragonal phases [26]. The magnetic structure of BNT is also very complex and completely different from BNT-PT [17]. At the applied magnetic field of lower strength, BNT exhibits an antiferromagnetic transition around 58 K. On increasing the field strength up to 7T, it shows a ferromagnetic like phase transition [17]. However, the BNT-PT solid solution is ferromagnetic in nature [16]. Ferromagnetic character increases with increasing BNT concentration within the composition range $0.80 \leq x < 1.0$. It has been proposed by Hu et al [16] that the ferromagnetism is linked with the super-exchange interaction in BNT-PT solid solution. Further increasing BNT content reduces the ferromagnetic nature due to the increasing a (=b) cell parameter of the tetragonal unit cell which weakens the super-exchange interaction. Ferromagnetic nature also decreases with increasing temperature due to the elongation of a (=b) tetragonal cell parameter [16]. Thus, BNT-PT is a multiferroic material with both ferroelectric and ferromagnetic ordering in the pure perovskite phase.

In recent years, large number of papers have been published on BNT-PT due to its high piezoelectric response and multiferroic character [16,24-31]. However, detailed Rietveld structural analysis has not been done to determine the structure of the MPB phase. Zhang et al. [26] have reported the presence of MPB in the composition range 0.46-0.48 but comprehensive structural study has not been done. Choi et al have [24] reported the phase diagram of BNT-PT using the results of calorimetric and dielectric studies without systematic structural studies. Kang et al. [29] have shown that the structure of BNT-PT across MPB is affected by the compositional fluctuations also. Modification of the crystal structure of BNT-PT is observed by the addition of excess amount of PbO, NiO, $TiO_2$ and $Bi_2O_3$ during synthesis. Doping of 1 to 3 mole % of PbO, NiO and $Bi_2O_3$ transforms the MPB phase into tetragonal phase but the effect is less prominent after the doping of 1 to 5 mole% $TiO_2$ [29]. Further, doping of small amount of Sr at Pb-site improves the problem of leakage current and reduces the loss [25]. Currently, BNT-PT based ternary systems such as (1-x-y)$PbTiO_3$-$xBi(Ni_{1/2}Ti_{1/2})O_3$-$yBiScO_3$ [31], (1-x-y)$PbTiO_3$-$xBi(Ni_{1/2}Ti_{1/2})O_3$-$yBiFeO_3$ [32] and $PbTiO_3$-$Bi(Ni_{1/2}Ti_{1/2})O_3$-$PbZrO_3$ [33] have been reported with excellent ferroelectric and magnetic properties.

To the best of our knowledge, systematic structural studies have not been done on the BNT-PT solid solution to determine the phase coexistence region and structure of the MPB phase. The preliminary studies by earlier workers have reported the coexistence of rhombohedral and tetragonal phases in the MPB region [26]. However, the existence of a monoclinic phase in the MPB region similar to other MPB systems has not been checked by the Rietveld method of structural analysis. In the present work, we have refined the structure of BNT-PT across MPB in the composition range of $0.35 \leq x \leq 0.55$ at close compositional interval using Rietveld method. We find that a monoclinic phase (Pm) coexists with the tetragonal phase (P4mm) in the MPB region. We have also determined precisely the stability region of various crystallographic phases at room temperature across the MPB.

**II. EXPERIMENTAL DETAILS**

Samples used in the present work were prepared by conventional solid state ceramic method. AR grade $Bi_2O_3$ (HIMEDIA, ≥99.5%), $NiCO_3.2Ni(OH)_2$ (QUALIGENS), $TiO_2$ (HIMEDIA, ≥99%), PbO (HIMEDIA, 98%) were used as raw materials. Pure NiO from $NiCO_3.2Ni(OH)_2$ was obtained after heating it at 550 $^0$C for 1 h. Mixing of various ingredients in stoichiometric proportions was carried out for 6 h using a ball mill (Retsch, Germany) with zirconia jars and zirconia balls. AR grade acetone was used as the mixing media. Heat treatments for calcination were carried out at 850 $^0$C in alumina crucibles using a muffle furnace. Multistep calcinations at 850 $^0$C were carried out to prepare the single phase BNT-PT. Cold compaction of calcined powders was done using a steel die of 12 mm diameter and a uniaxial hydraulic press at an optimized load of 65 kN. 2% polyvinyl alcohol solution in water was used as binder. The green pellets were kept at 500 $^0$C for 10 h to burn out the binder and then sintered at 950 $^0$C for 3 h in closed alumina crucible. The calcined powder of the same composition was kept inside the closed crucible as sacrificial powder for preventing the loss of volatile Bi and Pb oxides during sintering. For x-ray diffraction (XRD) measurements, sintered pellets were crushed into fine powder and then annealed at 500 $^0$C for 10 h to remove the strains introduced during crushing. XRD measurement was carried out using an 18 kW rotating Cu-anode based Rigaku (Japan) powder diffractometer operating in the Bragg-Brentano geometry and fitted with a curved crystal graphite monochromator. Rietveld structure refinements were carried out using FULLPROF suite [34]. The compositions close to MPB exhibited anisotropic peak broadening. The anisotropic peak shape function suggested by Stephens and also incorporated in the Fullprof program was used to refine the structure [34]. In the refinement process the background was modelled using 5-th order polynomial while the peak shapes were modeled by pseudo-Voigt function. Occupancy parameters of all the ions were fixed at the nominal composition during refinement. Zero correction, scale factor, background, half width parameters (U, V and W), lattice parameters, positional coordinates and thermal parameters were varied during the refinement. Anisotropic thermal parameters were used for *A*-site cations (Pb/Bi) for all the compositions except for the cubic compositions with x≤0.40. In the cubic phase with $Pm\bar{3}m$ space group, the occupancy of Bi/Pb ions were fixed at 1(a) sites (0, 0, 0), Ti/Ni ions at 1(b) sites (1/2, 1/2, 1/2) and oxygen ions at 3(c) sites (1/2, 1/2, 0). For the refinement of the rhombohedral structure with space group $R\bar{3}m$, we used the hexagonal unit cell having lattice parameters $a_H=b_H=\sqrt{2}a_R$ and $c_H=\sqrt{3}a_R$, where $a_R$ corresponds to rhombohedral lattice parameter. To fix the origin for the rhombohedral structure, the z-coordinate of O ion was fixed at 1/6. In the tetragonal phase with P4mm space group, the occupancy of Bi/Pb ions were fixed at 1(a) sites (0, 0, z) Ti/Ni and $O_I$ in 1(b) sites at (1/2, 1/2, z), and $O_{II}$ in 2(c) sites at (1/2, 0, *z*). For the monoclinic phase with space group Cm, the occupancy of Bi/Pb, Ti/Ni and $O_I$ were fixed in 2(a) sites at (x, 0, z) and $O_{II}$ in 4(b) sites at (x, y, z). For the monoclinic phase with space group Pm, the occupancy of Bi/Pb and $O_I$ ions were fixed in 1(a) sites at (x, 0, *z*), Ti/Ni, $O_{II}$ and $O_{III}$ in 1(b) sites at (x, 1/2, *z*).

**III. RESULTS AND DISCUSSION**

Fig.1 shows the powder XRD profiles of the (200), (220) and (222) pseudocubic reflections for BNT-PT solid solution in the composition range x=0.35 to 0.55. It is evident from Fig.1 that there is no splitting in the XRD profiles for the compositions with x≤0.40 and all the XRD profiles are singlet which suggest the cubic structure for these compositions. For the composition with x=0.41, a tail appears towards the lower 2θ side for (200) pseudocubic reflection and higher 2θ side for (220) pseudocubic reflection. The (222) pseudocubic profile is also significantly broader than that for x=0.40 and an asymmetry is observed on the lower 2θ side. This suggests that the structure of the composition with x=0.41 is neither tetragonal nor rhombohedral but either coexistence of these two phases or a monoclinic/orthorhombic reported in several MPB systems, recently [2-8]. If the dominant phase has a monoclinic structure then the coexistence of monoclinic phase with tetragonal phase may be expected in the PT rich compositions. For x≥0.50, the (200) and (220) profiles appear as doublet and (222) profile as a singlet, which suggests that the dominant phase for these compositions has tetragonal structure. It can be seen from Fig.1 that the (002) profile is significantly broader than that of (200) for the compositions with x=0.49 and 0.50. The anisotropic peak shape function suggested by Stephens [34] is used to get the satisfactory fit for the Rietveld structure refinement of compositions close to the MPB. We have also introduced the anisotropic peak shape function during the Rietveld structure refinement of BNT-PT solid solution.

Fig.2 depicts the variation of the FWHM for (222) pseudocubic profile as a function of composition (x) in the composition range $0.35 \leq x \leq 0.55$. The (222) profile should be a singlet in the tetragonal and cubic structures, whereas more than one reflection should be appear for phase coexistence or monoclinic structure. As can be seen from Fig.2, the FWHM is significantly enhanced for the composition range $0.41 \leq x < 0.49$, while it decreases on the either side. This clearly suggests the presence of lower symmetry phase or a phase coexistence region for these compositions [35]. The phase coexistence region reported by earlier researchers is in agreement with these results [16,24,26]. We have found that the large broadening of the (222) profile transform in to clear splitting corresponding to monoclinic phase on the application of electric field and the results will be published elsewhere. Very recently Tutuncu et al [36] have also reported significant modification in the structure of 0.55BNT-0.45PT, after application of electric field. On the application of electric field the MPB phase transforms in to a mix phase of tetragonal and pseudocubic structures [36]. However, Tutuncu et al [36] did not observe monoclinic structure in there samples perhaps due to in-situ measurements. In the subsequent sections, we present the results of Rietveld analysis to unambiguously determine the structure of BNT-PT ceramics across the MPB.

**A. Cubic structure with space group $Pm\bar{3}m$ (x≤0.40)**

Rietveld fit for the BNT-PT composition with x=0.40 using cubic structure (space group $Pm\bar{3}m$) is given in Fig.3. The inset shows the goodness of fit for the (111) profile. It is clear from the inset to Fig.3 that (111) profile is a singlet and no asymmetry or splitting is observed in lower 2-theta side, as observed in the rhombohedral phase of other MPB systems like PZT, PMN-PT etc [2-8]. Very high isotropic thermal parameters were observed for the *A*-site (~ 4.0(7) Å$^2$), *B*-site (~ 2.3(1) Å$^2$) cations and O anions (~ 6(4) Å$^2$). The large values of thermal parameters suggest that the system is highly disordered. The Rietveld structure refinement for the composition with x=0.35 also confirms the cubic structure with space group $Pm\bar{3}m$.

**B. Monoclinic structure with space group Pm (0.41≤x<0.43)**

In order to determine the structure for the composition range x=0.41 to 0.43, we have carried out the Rietveld analysis considering different plausible space groups viz. Cm, Pm and $R\bar{3}m$. The Rietveld fits for the (200), (220) and (222) pseudocubic reflections for the composition with x=0.41 are shown in Fig.4. The best fit between observed and calculated profiles is obtained for the space group Pm. Although, the satisfactory fit is found for the (200) and (220) profiles using space group Cm but it is poor for the (222) pseudocubic profile, as can be seen from the difference plot. It is further corroborated with the lowest value of $\chi^2$ obtained for the Pm space group in the composition range 0.41-0.43. In contrast, the rhombohedral phase with space group ($R\bar{3}m$) gives very poor fit for all the profiles. Consideration of minor coexisting tetragonal phase along with the monoclinic 'Pm' phase improves further the Rietveld fit and gives lower $\chi^2$. Thus, the structure is predominately monoclinic with space group Pm for the composition range x=0.41-0.43. The refined structural parameters and agreement factors are listed in Table. I for the composition range 0.41≤x<0.43 using the monoclinic structure (Pm).

**C. Phase co-existence region with Pm and P4mm space groups (0.43≤x≤0.49)**

In addition to Fig.2 and the discussion made earlier, the FWHM for the (222) pseudocubic profile is significantly enhanced for the composition range 0.43≤x≤0.49, suggesting phase coexistence region. In order to determine the structure of the phase coexistence region, we carried out the Rietveld analysis of the powder XRD data using various plausible structures. The Rietveld fit for (200), (220) and (222) pseudocubic reflections using (Pm+P4mm), (Cm+P4mm) and ($R\bar{3}m$+P4mm) structural models are shown in Fig.5 for x=0.46. It is evident from Fig.5 that the Rietveld fit for the $R\bar{3}m$ phase coexisting with the P4mm phase is very poor. The fit for the (Cm+P4mm) is better than ($R\bar{3}m$+P4mm) model, but significantly inferior than that obtained for (Pm+P4mm) model. The coexistence of (Pm+P4mm) provides the lowest value of $\chi^2$. Thus, the structure of BNT-PT consists of coexisting monoclinic (Pm space group) and tetragonal (P4mm space group) phases for the composition range 0.43≤x≤0.49 rather than ($R\bar{3}m$+P4mm) as reported by earlier authors [24, 26]. The refined structural parameters and the agreement factors are listed in Table I for the composition range 0.43≤x≤0.49 using the monoclinic structure (Pm) as the major phase and tetragonal (P4mm) as minor phase.

**D. Tetragonal structure with space group P4mm (x≥0.50)**

The Rietveld fit for the composition with x=0.55 using tetragonal structure with space group P4mm is shown in Fig.6. The fit is quite satisfactory which suggests that the structure is tetragonal (P4mm) for this composition. As observed for the tetragonal compositions close to the MPB for other MPB ceramics like PZT and PMN-PT, consideration of small coexisting monoclinic phase further improves the fit. The refined structural parameters and the agreement factors are listed in Table I for x≥0.50 using the tetragonal structure (P4mm) as major phase.

**E. Variation of lattice parameters with composition**

Fig.7 depicts the variation of lattice parameters with composition for BNT-PT solid solution. For cubic compositions, the lattice parameter 'a' has slightly decreasing trend with increasing PT concentration and corresponds to monoclinic $a_m$ parameter. Monoclinic cell parameter $c_m$ increases while $a_m$ and $b_m$ decrease with increasing PT concentration inside the MPB region. The tetragonal cell parameter $c_t$ increases while $a_t$ decreases with increasing PT concentration. The monoclinic angle (β) increases with increasing the major monoclinic phase (Pm) and start decreasing with increasing tetragonal phase fraction outside the MPB region. The variation of unit cell volume with composition (x) is shown in Fig.8. Increasing trend of cell volume is observed for the monoclinic and tetragonal phase regions possibly due to the development of the ferroelectric order. The enhancement in the unit cell volume on increasing the PT concentration is also expected due to increasing ionic radii of the cations at the *A*-sites. The average size of *B*-site cation in BNT ($r_{Bav}$= 0.6475 Å) is close to the size of *B*-site cation ($r_{Ti4+}$=0.605 Å) in PT but the size of the 12-coordinated $Pb^{2+}$ (1.49 Å) is significantly larger than $Bi^{3+}$ (1.36 Å) [37]. Cubic unit cell volume unexpectedly decreases with increasing 'x' even though it should increase due to increasing ionic radii of the cations at the *A*-sites.

**F. Variation of the (*B-O*) bond lengths with composition**

First-principles density functional theory (DFT) calculations by Qi et al. [38] suggest that the tetragonality (c/a) is strongly coupled with the *B*-site cation displacement in Bi/Pb-based solid solutions and hence the nature of *B-O* bonding in $ABO_3$ perovskite. Variations in the bond lengths for the *B*-site (Ni/Ti) cations with apical and planer oxygen anions as the function of composition are shown in Figs.9(a,b,c). The (*B-O*) bond lengths calculated on the basis of Shannon-Prewitt ionic radii are shown by solid (blue) lines in the above figures. As can be seen from Fig.9(a), for the cubic compositions (x≤0.40), where the Wyckoff positions are alike for all O-ions, the calculated bond length is nearly equal to the experimental value, suggesting the ionic nature of the Ni/Ti-O bonds. In this case, we get six Ni/Ti-O bonds of same length. With increasing the PT content, the Ni/Ti-O bond lengths splitted in to five unequal parts, as space group symmetry changes from cubic to monoclinic. Due to three different types of Wyckoff positions occupied by O-ions (see Table.I), we get two Ni/Ti-$O_I$ bonds of same length, two Ni/Ti-$O_{II}$ and Ni/Ti-$O_{III}$ bonds of unequal length as shown in Fig.9(b). This compositional region (0.41≤x≤0.49) corresponds to MPB region where the monoclinic phase is the major phase. Fig.9(c) depicts the variation of Ni/Ti-O bonds for tetragonal compositions with x≥0.50. Due to the availability of two different types of Wyckoff positions for O-ions in tetragonal (P4mm) structure, four Ni/Ti-$O_{II}$ bonds are same in length while two Ni/Ti-$O_I$ bond lengths are unequal.

In order to understand the nature of Ni/Ti-O bonding in BNT-PT solid solution, we compared the Ti-O bonds in $PbTiO_3$ with Ni/Ti-O bonds for various compositions of BNT-PT. Shirane et al. [39] have reported three types of Ti-O bonds (Ti-$O_{Ia}$=1.78 Å, Ti-$O_{Ib}$=2.38 Å and Ti-$O_{II}$=1.98 Å) in tetragonal $PbTiO_3$. The Ni/Ti-O bond lengths obtained by Rietveld refinement for BNT-PT come out to be 1.9836(1), 1.9833(1) and 1.9825(1) for the cubic compositions with x=0.35, 0.38 and 0.40 respectively, which are comparable to the Ti-$O_{II}$ bond length in $PbTiO_3$. The Ni/Ti-O bond lengths are listed in Table.II for the MPB compositions in the composition range 0.41≤x≤0.49 with major phase as monoclinic (Pm) structure. It is evident from Table.II that Ni/Ti-$O_I$, Ni/Ti-$O_{IIa}$ and Ni/Ti-$O_{IIb}$ bonds are comparable to Ti-$O_{II}$ bond length in $PbTiO_3$ and reveal the ionic nature. Ni/Ti-$O_{IIIa}$ bond lengths in BNT-PT are comparable to the Ti-$O_{Ib}$ bond length in $PbTiO_3$. However Ni/Ti-$O_{IIIb}$ bond lengths are shorter than the bond lengths in $PbTiO_3$, as reported by Shirane et al. [39] and calculated using Shannon-Prewitt radii [37]. If we consider the ionic nature of the bonds then all bond lengths should have been equal, as has been observed in the case of cubic phase. This clearly indicates the partial covalent nature of Ni/Ti-$O_{IIIb}$ bonds in BNT-PT solid solution. The Ni/Ti-O bond lengths obtained by Rietveld refinement of BNT-PT for the tetragonal composition with x≥0.50 are listed in Table.III. It is evident from the Table.III that Ni/Ti-O$_{II}$ bond lengths are almost equal to the Ti-O$_{II}$ bond length in PbTiO$_3$. Ni/Ti-O$_{Ia}$ and Ni/Ti-O$_{Ib}$ bond lengths in BNT-PT are slightly lower but comparable to Ti-O$_I$ bond length in PbTiO$_3$. The above results suggest that the tetragonality in BNT-PT can be described in the similar manner as explained in the PbTiO$_3$. It is reported by Cohen [40] that in the pure PbTiO$_3$, the hybridization between Ti$^{4+}$(3d$^o$) and O$^{2-}$(2p) soften the short range repulsion and results in ferroelectricity.

## G. DISCUSSION

The results discussed in the foregoing sections confirm that the structure of the morphotropic phase for BNT-PT ceramic is monoclinic with space group Pm. In contrast, the symmetry of the MPB phase in multiferroic PFN-PT solid solution is monoclinic in the space group Cm [41] whereas the multiferroic BF-PT solid solution exhibit monoclinic phase in the Cc space group [10]. The phase transition from tetragonal to rhombohedral phase is forbidden by the group-subgroup symmetry relationship and therefore requires the presence of some other phase (monoclinic) mediating between them. Stability of monoclinic phases in ferroelectric perovskites may be explained by considering eighth coefficient free energy in the expansion of Landau-Devonshire (LD) theory of ferroelectric homogeneous state [42]. Ever since the discovery of the MPB in the PZT and similar solid solutions, it is believed that the coexistence of the rhombohedral and tetragonal phases, makes available total 14 (6 for tetragonal and 8 for rhombohedral) orientational directions for the polarization, thereby maximizing the piezoelectric response [2]. Later on Noheda et al [8] discovered a monoclinic phase in the PZT system followed by the discovery of monoclinic phases in several other MPB solid solutions such as PMN-PT [4-5], PZN-PT [6], PSN-PT [43], BS-PT [44] etc. First principles calculations by Fu and Cohen [9] suggest that rotation of polarization vector in planes corresponding to the monoclinic phase leads to ultra high electromechanical response in the MPB region. The polarization vector has more freedom to rotate in the monoclinic phase in comparison to the coexisting rhombohedral and tetragonal phases. Recently it has been observed experimentally that significant domain wall and interphase boundary motion occurs in the MPB region and domain wall motion within the monoclinic phase has strong contribution to the electric field induced strain/piezoelectric response [45].

Jin et al [46] have reported that conformal miniaturization of stress accommodating tetragonal/rhombohedral domains under the condition of low domain wall energy density can mimic the monoclinic symmetry as observed in the diffraction experiments. It is reported that for very low values of domain-wall energies the system transforms into a mixed adaptive state which is inhomogeneous on the nanoscale and homogeneous on the macroscale [46]. Phase identification is usually carried out by the x-ray or neutron diffraction experiments. Coherence length of these radiations is much larger than the conformally miniaturized nanodomains. Therefore, the resultant crystal structure identified by the intensity of diffraction profile provides the average structure of materials. As a consequence, the average diffraction profile of phase-coexisting nanoscale lamellar domains of rhombohedral and tetragonal phases (nanotwins) seems to be monoclinic phase in the diffraction profiles [46]. A series of papers by Wang and co-workers [46-47] have also been shown that the monoclinic phase reported for the MPB can be expressed as conformally miniaturized stress accommodating tetragonal/rhombohedral domains. However, if one accepts these explanations then temperature dependence of the physical properties like elastic modulus and dielectric response should not show any anomaly corresponding to the phase transition from the monoclinic phase to the tetragonal phase [48-49]. Filho et al [50] have reported that the tetragonal to monoclinic phase transition in $PbZr_{0.52}Ti_{0.48}O_3$ is also accompanied by the appearance of new Raman mode. If monoclinic symmetry is resulting due to miniaturization of tetragonal domains then one should not get any new Raman mode in the Raman spectra. Schonau et al [51] have performed a detailed synchrotron x-ray diffraction and transmission electron microscopic (TEM) studies on several compositions of PZT to correlate domain structure and appearance of monoclinic phase around MPB. The TEM studies reveal the miniaturization of the tetragonal domains while approaching to the MPB compositions. However due to experimental limitations, these workers could not determine the symmetry of the miniaturized nanodomains. To reconcile all these observations with the appearance of the new Raman mode and anomalies in the temperature dependence of the physical properties corresponding to the tetragonal to monoclinic phase transition, we propose following model. As observed experimentally by Schonau et al [51] in TEM studies and also reported by Jin et al [46] and Wang et al [47], one cannot disagree with the presence of the conformally miniaturized domains. We believe that the conformal miniaturization of the rhombohedral/tetragonal domains in the compositions around MPB are responsible for the anisotropic broadening of the diffraction profiles of the rhombohedral and tetragonal phases, as one approaches the MPB. We observed significant anisotropic peak broadening in our BNT-PT samples for the compositions close to MPB. It is well known for the MPB systems like PZT [3,49] and PMN-PT [48] that anisotropic broadening of the diffraction profiles is observed for the rhombohedral as well as tetragonal compositions close to the MPB. Consideration of monoclinic symmetry for the so-called rhombohedral compositions gives a better fit for the diffraction profiles in the Rietveld structure refinement. Such compositions do not show tetragonal to monoclinic transition with temperature. However, if the symmetry is indeed lowered to the monoclinic structure for the MPB phase, one observes tetragonal to monoclinic phase transition. It has been found in the MPB ceramics like PZT and PMN-PT that the tetragonal and rhombohedral compositions do not exhibit peak broadening of the diffraction profiles away from the MPB where conformal miniaturization of domains is absent. As one approaches to MPB, gradually enhancing anisotropic peak broadening is observed for both the tetragonal and rhombohedral structures which eventually transforms to the monoclinic symmetry for the MPB compositions. This picture is also consistent with the results reported by Glazer et al [52] where they have shown that in the PZT system, the monoclinic character is present across the MPB with short range monoclinic ordering in the rhombohedral and tetragonal compositions which becomes long ranged in the MPB region. More theoretical and experimental investigations will be needed to further confirm this picture.

## G. CONCLUSION

To summarize, we have shown that the structure of the MPB phase in BNT-PT solid solution is monoclinic in the Pm space group. The structure of the compositions with $x \leq 0.40$ is cubic in the space group $Pm\bar{3}m$ and tetragonal in the space group P4mm for the compositions with $x \geq 0.50$. The MPB region lies in the composition range $0.41 \leq x \leq 0.49$. The composition with x=0.41, 0.42 and 0.43 is nearly single phase monoclinic while the other compositions in the MPB region exhibit coexistence of minority tetragonal phase with space group P4mm. The cubic phase has predominantly ionic nature of *B-O* (Ni/Ti-O) bonds while partial covalent character is observed for the other crystallographic phases. The correlation of anisotropic peak broadening and conformal miniaturization of rhombohedral/tetragonal domains in the vicinity of MPB with the appearance of monoclinic phase is also discussed.

## ACKNOWLEDGEMENTS

Authors are thankful to Professor Dhananjai Pandey, School of Materials Science and Technology, IIT (BHU) Varanasi, India for extending laboratory facilities. R.P acknowledges

University Grant Commission (UGC), India for the financial support as senior research fellow (SRF).

**FIGURE CAPTIONS:**

**Fig.1** Pseudocubic (200), (220) and (222) XRD profiles of $(1-x)Bi(Ni_{1/2}Ti_{1/2})O_3-xPbTiO_3$ solid solution for the compositions with x=0.35, 0.40, 0.41, 0.43, 0.45, 0.46, 0.47, 0.49, 0.50 and 0.55 sintered at 950 $^0$C. The arrows mark the peak positions for tetragonal (T) and monoclinic (M) phases.

**Fig.2** Variation of full width at half maxima (FWHM) of (222) pseudocubic reflection for $(1-x)Bi(Ni_{1/2}Ti_{1/2})O_3-xPbTiO_3$ with x=0.35, 0.38, 0.40, 0.41, 0.43, 0.45, 0.46, 0.47, 0.49, 0.50, 0.52 and 0.55.

**Fig.3** Observed (dots), calculated (continuous line) and difference (continuous bottom line) profiles for $(1-x)Bi(Ni_{1/2}Ti_{1/2})O_3-xPbTiO_3$ with x=0.40 obtained after Rietveld analysis of the powder XRD data using cubic (Pm$\bar{3}$m) structure. The vertical tick-marks above the difference plot show the peak positions. The inset illustrates the goodness of fit.

**Fig.4** Observed (dots), calculated (continuous line) and difference (continuous bottom line) profiles for (1-x)Bi(Ni$_{1/2}$Ti$_{1/2}$)O$_3$-xPbTiO$_3$ with x=0.41 obtained after Rietveld analysis of the powder XRD data using monoclinic (Pm), monoclinic (Cm) and Rhombohedral (R$\bar{3}$m) structures. The vertical tick-marks above the difference plot show the peak positions.

**Fig.5** Observed (dots), calculated (continuous line) and difference (continuous bottom line) profiles for (1-x)Bi(Ni$_{1/2}$Ti$_{1/2}$)O$_3$-xPbTiO$_3$ with x=0.46 obtained after Rietveld analysis of the powder XRD data using (Cm+P4mm), (Pm+P4mm) and (R$\bar{3}$m+P4mm) structures. The vertical tick-marks above the difference plot show the peak positions.

**Fig.6** Observed (dots), calculated (continuous line) and difference (continuous bottom line) profiles for (1-x)Bi(Ni$_{1/2}$Ti$_{1/2}$)O$_3$-xPbTiO$_3$ with x=0.55 obtained after Rietveld analysis of the powder XRD data using tetragonal (P4mm) structure. The vertical tick-marks above the difference plot show the peak positions. The inset illustrates the goodness of fit.

**Fig.7** Variation of lattice parameter with composition for (1-x)Bi(Ni$_{1/2}$Ti$_{1/2}$)O$_3$-xPbTiO$_3$ ceramics. The vertical dotted lines demarcate different phase fields.

**Fig.8** Variation of the unit cell volume (Å$^3$) with compositions for (1-x)Bi(Ni$_{1/2}$Ti$_{1/2}$)O$_3$-xPbTiO$_3$.

**Fig.9** Variations of Ni/Ti–O (i.e. B–O) bond lengths for (a) cubic (x≤0.40) (b) monoclinic (0.41≤x≤0.49), and (c) tetragonal (x≥0.50) phases with composition obtained after Rietveld refinement. Solid blue line shows the bond lengths calculated using Shannon-Prewitt ionic radii.

**Table.I:** Rietveld structural parameters for BNT-PT ceramics in the composition range 0.35≤x≤0.55 using various space groups.

| (x) | Ions | $x$ | $y$ | $z$ | Thermal parameters (Å$^2$) | Lattice Parameters (Å) |
|---|---|---|---|---|---|---|
| **0.41** | Bi/Pb | 0.0 | 0.0 | 0.0 | $\beta_{11}$=0.087(5), $\beta_{22}$=0.108(3), $\beta_{33}$=0.039(4) | a$_m$=3.9657(2) |
| (Pm) | Ni/Ti | 0.52(1) | 0.5 | 0.55(3) | Biso=0.04(3) | b$_m$=3.9533(3) |
|  | O$_I$ | 0.44(3) | 0.0 | 0.61(4) | Biso=0.15(2) | c$_m$=3.9815(6) |
|  | O$_{II}$ | 0.41(1) | 0.5 | 0.05(1) | Biso=0.10(3) | $\beta$=90.101(3)° |
|  | O$_{III}$ | -0.08(1) | 0.5 | 0.58(4) | Biso=0.12(1) |  |
| **0.42** | Bi/Pb | 0.0 | 0.0 | 0.0 | $\beta_{11}$=0.102(1), $\beta_{22}$=0.113(2), $\beta_{33}$=0.045(3) | $a_m$=3.9657(3) |
| (Pm) | Ni/Ti | 0.523(8) | 0.5 | 0.553(4) | Biso=0.07(3) | $b_m$=3.9519(3) |
| | $O_I$ | 0.44(2) | 0.0 | 0.61(3) | Biso=0.17(2) | $c_m$=3.9886(8) |
| | $O_{II}$ | 0.41(2) | 0.5 | 0.05(2) | Biso=0.10(3) | $\beta$=90.108(1)° |
| | $O_{III}$ | -0.05(3) | 0.5 | 0.59(1) | Biso=0.13(1) | |
| **0.43** | Bi/Pb | 0.0 | 0.0 | 0.0 | $\beta_{11}$=0.081(1), $\beta_{22}$=0.10(8), $\beta_{33}$=0.052(4) | $a_m$=3.9649(5), |
| (Pm) | Ni/Ti | 0.527(5) | 0.50 | 0.548(2) | Biso=0.07(3) | $b_m$=3.9507(3) |
| | $O_I$ | 0.46(2) | 0.0 | 0.61(8) | Biso=0.17(2) | $c_m$=3.9883(5) |
| | $O_{II}$ | 0.41(3) | 0.5 | 0.09(1) | Biso=0.10(3) | $\beta$=90.113(4)° |
| | $O_{III}$ | -0.02(2) | 0.5 | 0.56(3) | Biso=0.13(1) | |
| (P4mm) | Bi/Pb | 0.0 | 0.0 | 0.0 | $\beta_{11}$=$\beta_{22}$=0.030(2), $\beta_{33}$=0.011(2) | $a_t$=3.9410(3) |
| | Ni/Ti | 0.5 | 0.5 | 0.552(3) | Biso=0.15(1) | $c_t$=4.0283(5) |
| | $O_I$ | 0.5 | 0.5 | 0.13(1) | Biso=1.2(2) | |
| | $O_{II}$ | 0.5 | 0.0 | 0.651(7) | Biso=0.8(1) | |
| **0.44** | Bi/Pb | 0.0 | 0.0 | 0.0 | $\beta_{11}$=0.086(6), $\beta_{22}$=0.094(8), $\beta_{33}$=0.051(5) | $a_m$=3.9648(5) |
| (Pm) | Ni/Ti | 0.526(5) | 0.0 | 0.550(5) | Biso=0.04(3) | $b_m$=3.9493(3) |
| | $O_I$ | 0.45(3) | 0.0 | 0.60(3) | Biso=0.17(2) | $c_m$=3.9929(1) |
| | $O_{II}$ | 0.41(1) | 0.5 | 0.07(1) | Biso=0.10(3) | $\beta$=90.125(2)° |
| | $O_{III}$ | -0.07(1) | 0.5 | 0.60(3) | Biso=0.13(1) | |
| (P4mm) | Bi/Pb | 0.0 | 0.0 | 0.0 | $\beta_{11}$=$\beta_{22}$=0.039(1), $\beta_{33}$=0.011(2) | $a_t$=3.9400(2) |
| | Ni/Ti | 0.5 | 0.5 | 0.549(3) | Biso=0.15(1) | $c_t$=4.0285(4) |
| | $O_I$ | 0.5 | 0.5 | 0.123(7) | Biso=1.2(2) | |
| | $O_{II}$ | 0.5 | 0.0 | 0.626(4) | Biso=0.8(1) | |
| **0.45** | Bi/Pb | 0.0 | 0.0 | 0.0 | $\beta_{11}$=0.079(5), $\beta_{22}$=0.098(6), $\beta_{33}$=0.054(2) | $a_m$=3.9626(4) |
| (Pm) | Ni/Ti | 0.532(4) | 0.5 | 0.548(3) | Biso=0.04(3) | $b_m$=3.9470(3) |
| | $O_I$ | 0.44(1) | 0.0 | 0.62(1) | Biso=0.17(2) | $c_m$=3.9953(5) |
| | $O_{II}$ | 0.41(1) | 0.5 | 0.08(1) | Biso=0.10(3) | $\beta$=90.130(3)° |
| | $O_{III}$ | -0.06(3) | 0.5 | 0.59(2) | Biso=0.13(1) | |
| (P4mm) | Bi/Pb | 0.0 | 0.0 | 0.0 | $\beta_{11}$=$\beta_{22}$=0.035(1), $\beta_{33}$=0.021(2) | $a_t$=3.9381(2) |
| | Ni/Ti | 0.5 | 0.5 | 0.544(3) | Biso=0.19(2) | $c_t$=4.0336(4) |
| | $O_I$ | 0.5 | 0.5 | 0.171(8) | Biso=1.3(1) | |
| | $O_{II}$ | 0.5 | 0.0 | 0.645(5) | Biso=0.9(1) | |
| **0.46** | Bi/Pb | 0.0 | 0.0 | 0.0 | $\beta_{11}$=0.101(8), $\beta_{22}$=0.113(1), $\beta_{33}$=0.057(6) | $a_m$=3.9663(6) |
| (Pm) | Ni/Ti | 0.532(8) | 0.5 | 0.555(6) | Biso=0.04(3) | $b_m$=3.9491(4) |
| | $O_I$ | 0.44(3) | 0.0 | 0.62(3) | Biso=0.17(2) | $c_m$=3.9977(8) |
| | $O_{II}$ | 0.41(1) | 0.5 | 0.05(1) | Biso=0.10(3) | $\beta$=90.139(1)° |
| | $O_{III}$ | -0.07(1) | 0.5 | 0.58(4) | Biso=0.13(1) | |
| (P4mm) | Bi/Pb | 0.0 | 0.0 | 0.0 | $\beta_{11}$=$\beta_{22}$=0.041(1), $\beta_{33}$=0.018(1) | $a_t$=3.9381(2) |
| | Ni/Ti | 0.5 | 0.5 | 0.542(2) | Biso=0.19(2) | $c_t$=4.0338(2) |
| | $O_I$ | 0.5 | 0.5 | 0.102(6) | Biso=1.3(1) | |
| | $O_{II}$ | 0.5 | 0.0 | 0.630(4) | $B_{iso}$=0.9(1) | |
| **0.47** | Bi/Pb | 0.0 | 0.0 | 0.0 | $\beta_{11}$=0.084(2),$\beta_{22}$=0.10(1),$\beta_{33}$=0.076(1) | $a_m$=3.9622(5) |
| (Pm) | Ni/Ti | 0.528(6) | 0.0 | 0.551(5) | $B_{iso}$=0.04(3) | $b_m$=3.9464(3) |
| | $O_I$ | 0.45(2) | 0.0 | 0.60(1) | $B_{iso}$=0.17(2) | $c_m$=3.9946(5) |
| | $O_{II}$ | 0.39(1) | 0.5 | 0.08(1) | $B_{iso}$=0.10(3) | $\beta$=90.142(2)° |
| | $O_{III}$ | -0.04(1) | 0.5 | 0.55(2) | $B_{iso}$=0.13(1) | |
| (P4mm) | Bi/Pb | 0.0 | 0.0 | 0.0 | $\beta_{11}=\beta_{22}$=0.038(3),$\beta_{33}$=0.012(1) | $a_t$=3.9376(2) |
| | Ni/Ti | 0.5 | 0.5 | 0.548(2) | $B_{iso}$=0.19(2) | $c_t$=4.0358(3) |
| | $O_I$ | 0.5 | 0.5 | 0.102(6) | $B_{iso}$=1.3(1) | |
| | $O_{II}$ | 0.5 | 0.0 | 0.639(3) | $B_{iso}$=0.9(1) | |
| **0.48** | Bi/Pb | 0.0 | 0.0 | 0.0 | $\beta_{11}$=0.068(2),$\beta_{22}$=0.13(1),$\beta_{33}$=0.056(7) | $a_m$=3.9649(3) |
| (Pm) | Ni/Ti | 0.541(9) | 0.5 | 0.554(8) | $B_{iso}$=0.04(3) | $b_m$=3.9464(7) |
| | $O_I$ | 0.44(3) | 0.0 | 0.63(1) | $B_{iso}$=0.15(2) | $c_m$=3.9983(1) |
| | $O_{II}$ | 0.41(0) | 0.5 | 0.05(2) | $B_{iso}$=0.10(3) | $\beta$=90.140(2)° |
| | $O_{III}$ | -0.06(5) | 0.5 | 0.56(3) | $B_{iso}$=0.12(2) | |
| (P4mm) | Bi/Pb | 0.0 | 0.0 | 0.0 | $\beta_{11}=\beta_{22}$=0.043(4),$\beta_{33}$=0.013(2) | $a_t$=3.9375(2) |
| | Ni/Ti | 0.5 | 0.5 | 0.538(3) | $B_{iso}$=0.19(2) | $c_t$=4.0361(4) |
| | $O_I$ | 0.5 | 0.5 | 0.135(6) | $B_{iso}$=1.2(1) | |
| | $O_{II}$ | 0.5 | 0.0 | 0.638(4) | $B_{iso}$=0.8(1) | |
| **0.49** | Bi/Pb | 0.0 | 0.0 | 0.0 | $\beta_{11}$=0.081(6),$\beta_{22}$=0.124(4),$\beta_{33}$=0.073(3) | $a_m$=3.9617(5) |
| (Pm) | Ni/Ti | 0.554(5) | 0.5 | 0.552(5) | $B_{iso}$=0.04(3) | $b_m$=3.9424(4) |
| | $O_I$ | 0.43(2) | 0.0 | 0.59(3) | $B_{iso}$=0.15(2) | $c_m$=3.9993(6) |
| | $O_{II}$ | 0.40(1) | 0.5 | 0.08(1) | $B_{iso}$=0.10(3) | $\beta$=90.120(3)° |
| | $O_{III}$ | -0.04(1) | 0.5 | 0.57(4) | $B_{iso}$=0.12(2) | |
| (P4mm) | Bi/Pb | 0.0 | 0.0 | 0.0 | $\beta_{11}=\beta_{22}$=0.041(5),$\beta_{33}$=0.025(1) | $a_t$=3.9368(1) |
| | Ni/Ti | 0.5 | 0.5 | 0.545(2) | $B_{iso}$=0.19(1) | $c_t$=4.0364(2) |
| | $O_I$ | 0.5 | 0.5 | 0.113(4) | $B_{iso}$=0.2(1) | |
| | $O_{II}$ | 0.5 | 0.5 | 0.634(2) | $B_{iso}$=0.2(0) | |
| **0.50** | Bi/Pb | 0.0 | 0.0 | 0.0 | $\beta_{11}=\beta_{22}$=0.041(5),$\beta_{33}$=0.028(1) | $a_t$=3.9351(3) |
| (P4mm) | Ni/Ti | 0.0 | 0.0 | 0.546(2) | $B_{iso}$=0.3(1) | $c_t$=4.0416(2) |
| | $O_I$ | 0.5 | 0.5 | 0.120(3) | $B_{iso}$=0.2(1) | |
| | $O_{II}$ | 0.5 | 0.0 | 0.629(3) | $B_{iso}$=0.2(0) | |
| **0.52** | Bi/Pb | 0.0 | 0.0 | 0.0 | $\beta_{11}=\beta_{22}$=0.014(5),$\beta_{33}$=0.024(1) | $a_t$=3.9324(1) |
| (P4mm) | Ni/Ti | 0.0 | 0.5 | 0.543(2) | $B_{iso}$=0.3(2) | $c_t$=4.0485(4) |
| | $O_I$ | 0.5 | 0.5 | 0.116(3) | $B_{iso}$=0.2(1) | |
| | $O_{II}$ | 0.5 | 0.0 | 0.626(3) | $B_{iso}$=0.2(0) | |

| 0.55   | Bi/Pb | 0.0 | 0.0 | 0.0      | $\beta_{11}=\beta_{22}=0.039(1)$, $\beta_{33}=0.027(1)$ |                  |
|--------|-------|-----|-----|----------|---------------------------------------------------------|------------------|
| (P4mm) | Ni/Ti | 0.0 | 0.5 | 0.547(2) | $B_{iso}=0.3(2)$                                        | $a_t=3.9322(1)$  |
|        | $O_I$ | 0.5 | 0.5 | 0.117(3) | $B_{iso}=0.2(1)$                                        | $c_t=4.0533(3)$  |
|        | $O_{II}$ | 0.5 | 0.0 | 0.624(2) | $B_{iso}=0.2(0)$                                     |                  |

**Table.II:** Ni/Ti-O bond lengths for BNT-PT solid solution obtained by using Rietveld refined structural parameters for the composition range 0.41≤x≤0.49 with monoclinic structure (Pm).

| (x)  | Ni/Ti-$O_I$ (Å) | Ni/Ti-$O_{IIa}$ (Å) | Ni/Ti-$O_{IIb}$ (Å) | Ni/Ti-$O_{IIIa}$ (Å) | Ni/Ti-$O_{IIIb}$ (Å) |
|------|-----------------|---------------------|---------------------|----------------------|----------------------|
| 0.41 | 2.011(5)        | 2.001(3)            | 2.063(1)            | 2.382(2)             | 1.590(2)             |
| 0.42 | 2.012(0)        | 1.994(1)            | 2.089(4)            | 2.379(2)             | 1.601(1)             |
| 0.43 | 2.013(2)        | 1.987(2)            | 2.095(4)            | 2.388(4)             | 1.561(6)             |
| 0.44 | 2.017(5)        | 1.961(2)            | 2.097(4)            | 2.454(8)             | 1.531(2)             |
| 0.45 | 2.020(5)        | 1.945(2)            | 2.110(1)            | 2.441(2)             | 1.541(1)             |
| 0.46 | 2.024(8)        | 1.932(0)            | 2.112(4)            | 2.398(6)             | 1.573(6)             |
| 0.47 | 2.030(2)        | 1.919(7)            | 2.092(3)            | 2.379(6)             | 1.649(3)             |
| 0.48 | 2.035(0)        | 1.961(5)            | 2.062(3)            | 2.403(0)             | 1.562(1)             |
| 0.49 | 2.028(1)        | 1.967(0)            | 2.120(4)            | 2.314(5)             | 1.596(6)             |

**Table.III:** Ni/Ti-O bond lengths for BNT-PT solid solution obtained by using Rietveld refined structural parameters for the tetragonal (P4mm) compositions with x=0.50, 0.52 and 0.55.

| (x)  | Ni/Ti-$O_{Ia}$ (Å) | Ni/Ti-$O_{Ib}$ (Å) | Ni/Ti-$O_{II}$ (Å) |
|------|--------------------|--------------------|--------------------|
| 0.50 | 1.722(1)           | 2.318(1)           | 1.995(5)           |
| 0.52 | 1.728(2)           | 2.319(3)           | 1.994(6)           |
| 0.55 | 1.735(3)           | 2.325(1)           | 1.992(1)           |

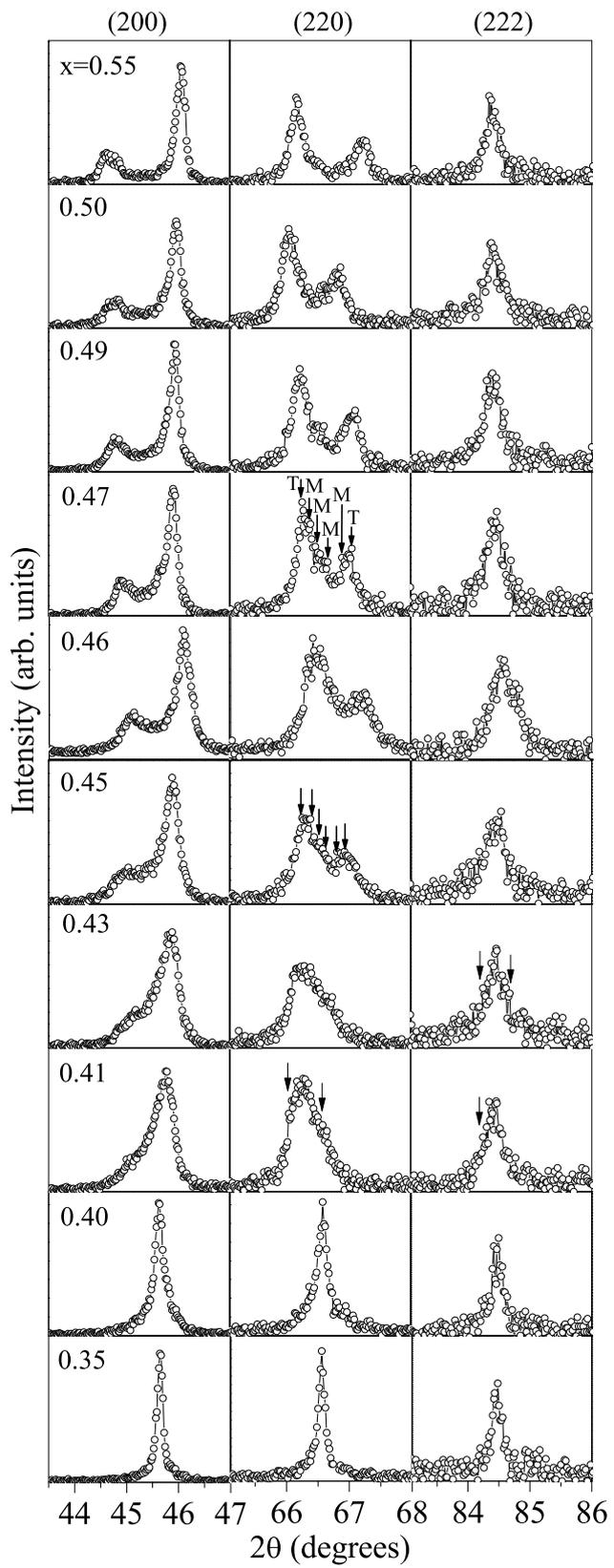

Fig.1

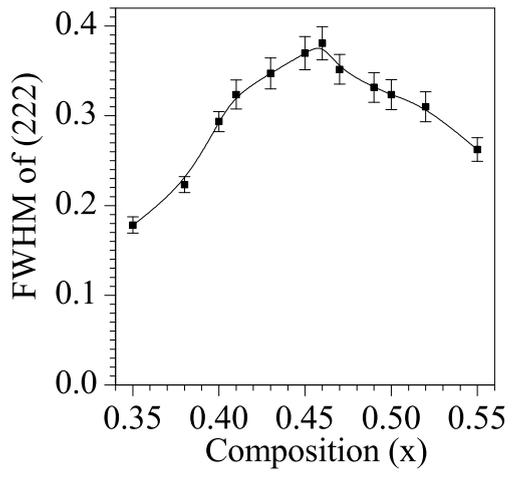

Fig.2

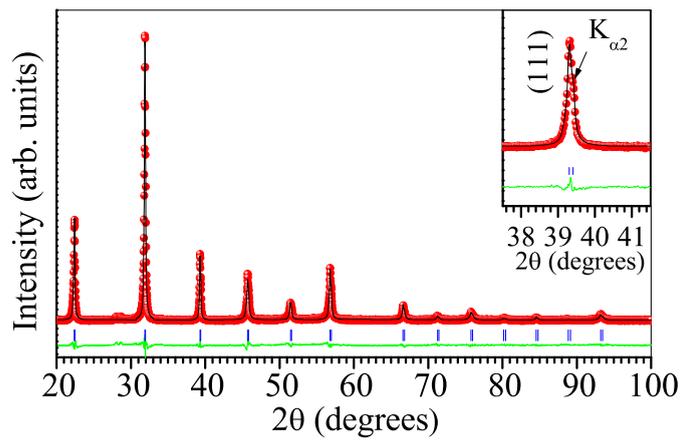

Fig.3

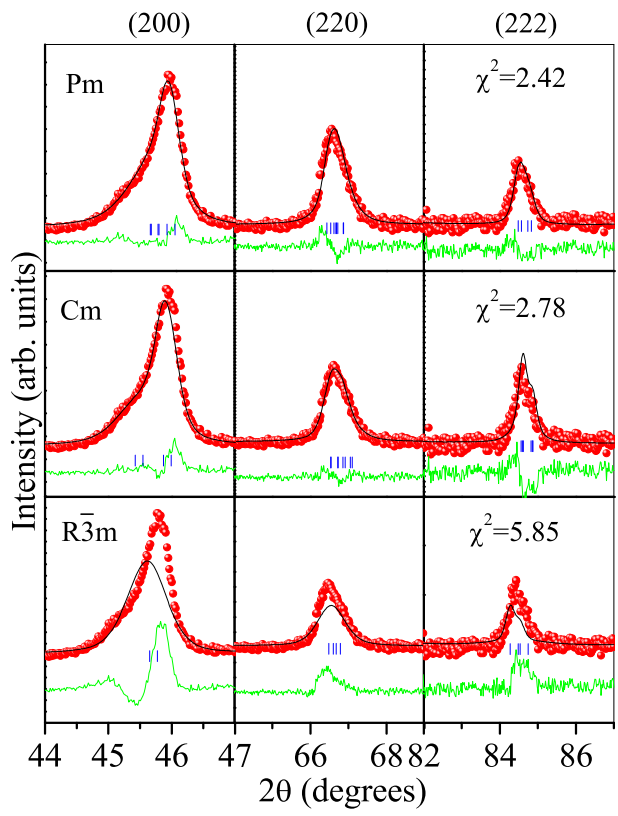

Fig.4

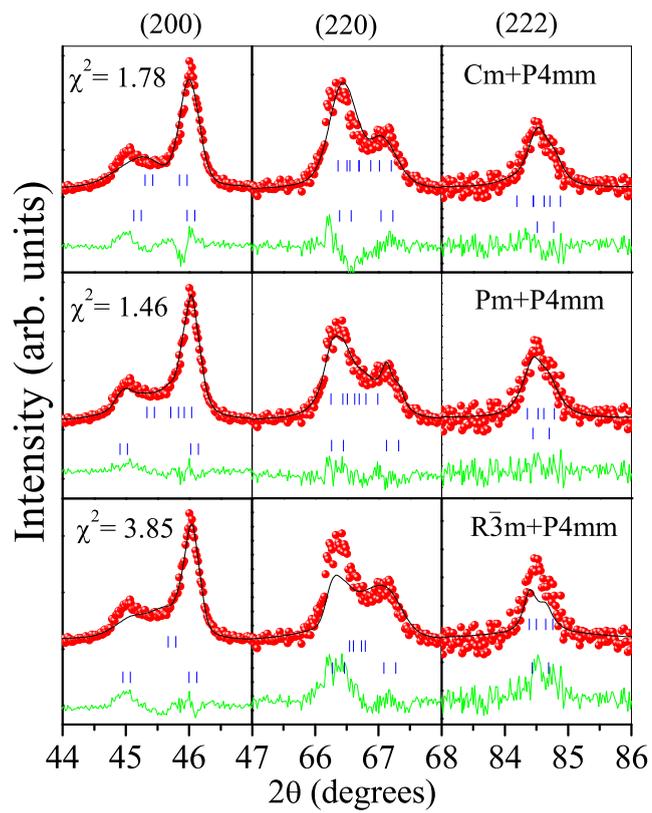

Fig.5

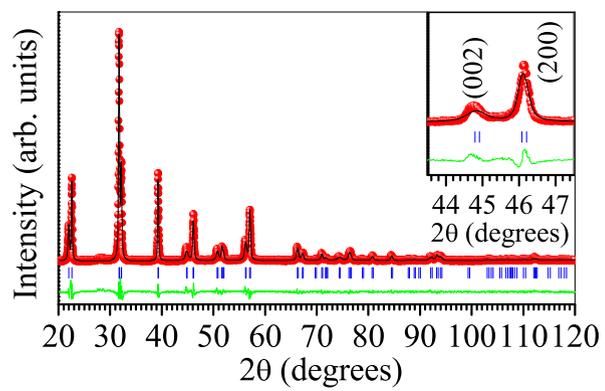

Fig.6

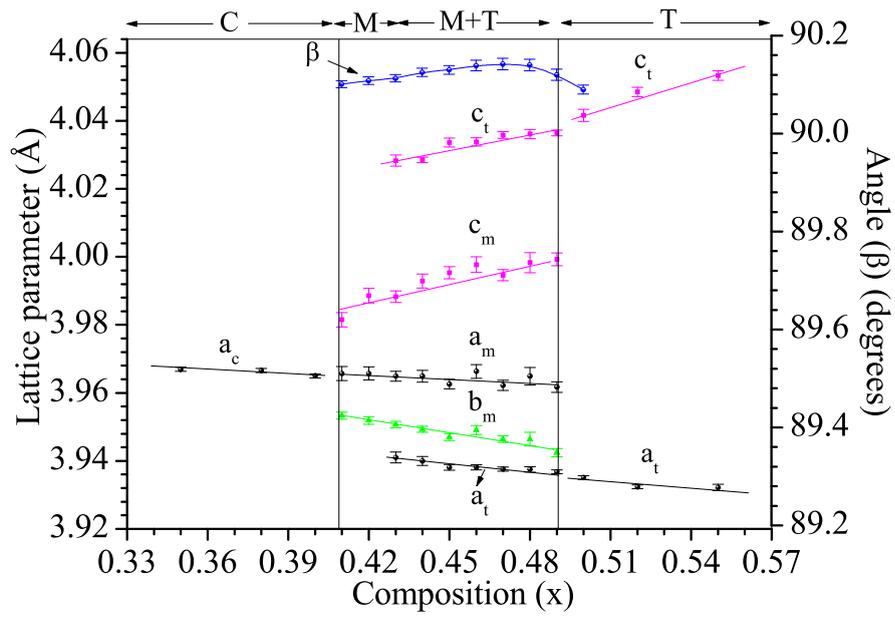

Fig.7

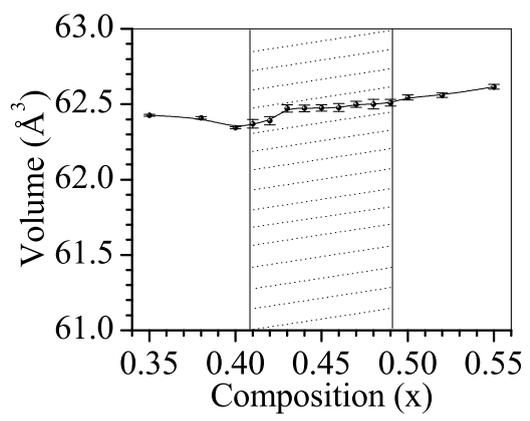

Fig.8

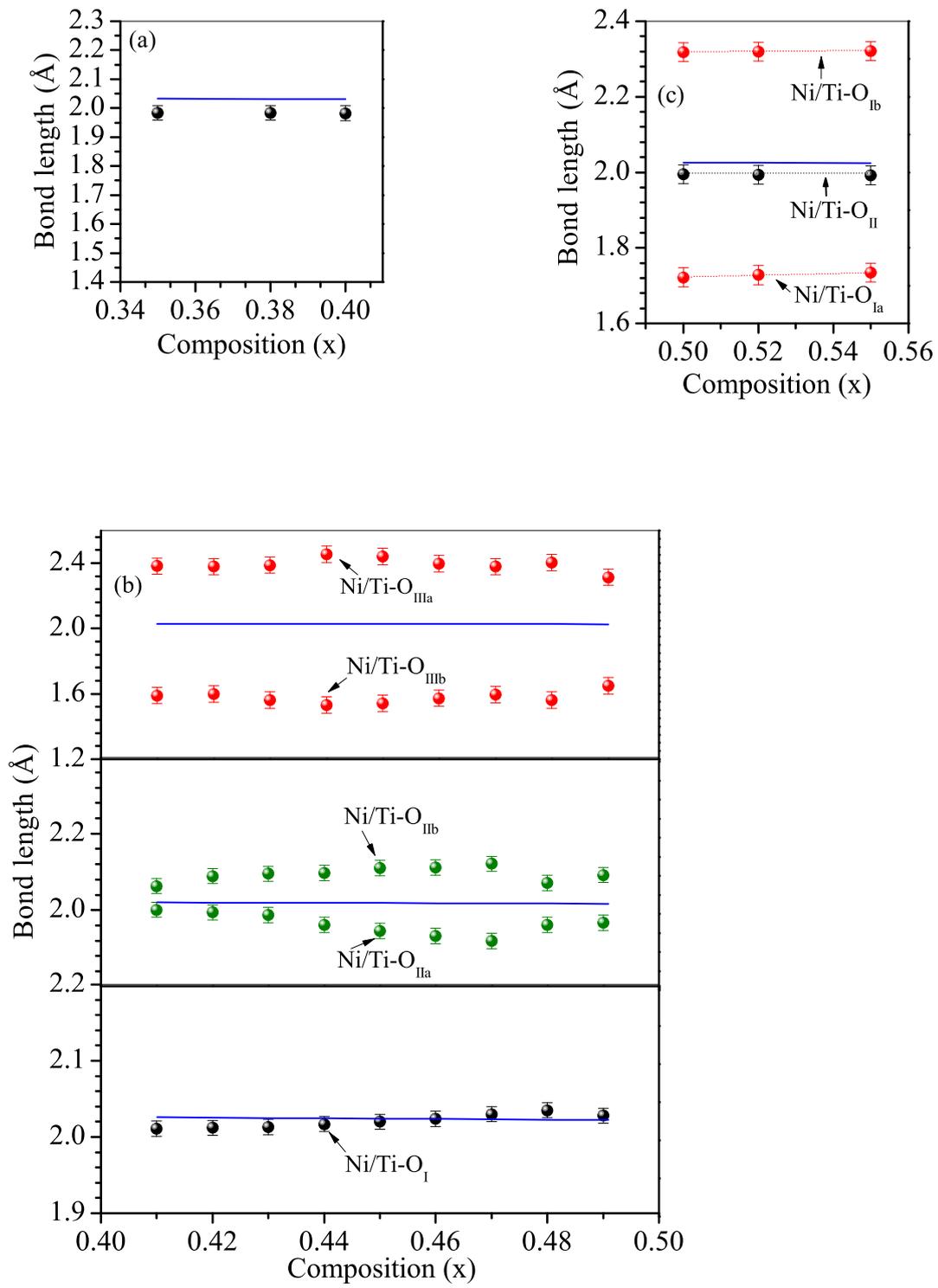

Fig.9